# The Interface Is Downstream

## Designing the Terms of Human-Agent Collaboration

An autobiographical design position grounded in the continued development of Alicia

Hector Ouilhet Olmos

***Position:*** *The terms of collaboration are set upstream. Attention governs what the teammate notices, evidence governs what it can claim, action boundaries govern what it can do, and learning governs what it carries forward.*

### Abstract

Before an agent responds or acts, much of the experience has already been designed. Memory and retrieval shape what it notices. Evidence rules shape what it may claim. Permissions shape what it can do. Learning rules shape what it carries into the next encounter.

The argument comes from Alicia, a personal agent I've built and used since January 2026. A fine-tuning pilot produced no defensible model-performance result. It exposed a provenance failure: Alicia repeated an interpretation from a retrieved synthesis, cited a source note credited by that synthesis, and left the synthesis out of the visible chain.

In a model-blind review of thirty-three citations, one reviewer judged that the retrieved intermediary supplied the claim in sixteen relayed citations and part of it in four. Five of twelve citations to directly retrieved targets lacked support in the target excerpt supplied for review. These judgments remain unadjudicated, and the packet is not public.

I call the shared setting a **humorphic environment**: a persistent computational setting that translates a human practice into software. The first Humorphism paper translated partnership. This paper translates the studio, the room where practice happens. The failure prompted an audit of attention, evidence, action, and learning, with a review artifact and available recourse for each. The test is whether the person can inspect and contest what shaped the teammate's behavior. The output is downstream. Correction, consent, and learning carry the collaborative interface back upstream.

### 1. The terms are set upstream

Alicia produced a Spanish response that looked well grounded. It cited notes on Antonio Damasio and Rick Rubin. Both notes existed in my vault. Neither had been retrieved for the prompt.

The document Alicia had received was one of its own syntheses, *Courage and vulnerability are the same gesture*. The synthesis described ideas from Damasio, Rubin, and one of my notes. It placed a source link beside each description. Alicia translated the claims, retained two links, and removed the synthesis that stood between the response and the named sources.

The full path was:

> response → retrieved Alicia synthesis → source credited by the synthesis

The interface rendered:

response → credited source

The claim had a supplier. Its history disappeared.

One generated sentence said, translated from Spanish, that courage requires vulnerability because a person must feel the body's fear signal rather than suppress it. The retrieved synthesis made that interpretation. The linked *Feeling & Knowing* note contained only book metadata and a broad Alicia-written bridge about feeling and thinking. It contained no passage about courage or somatic-marker patients.

The missing intermediary therefore mattered twice. It concealed the document that supplied the interpretation. It also made a sparse book note look like direct support for a stronger claim than the note contained.

A surface review could still find the response fluent, relevant, and useful. The citation resolved to a real file. The failure appeared only when I followed the claim backward through retrieval and authorship. By the time the words reached the interface, the consequential choices had already happened.

The terms of collaboration are set upstream. Attention governs what the teammate notices on the person's behalf. Evidence governs what it may claim and what the person can challenge. Action boundaries govern where initiative ends and consent begins. Learning governs which interactions alter the future relationship.

The output surface is downstream from those decisions. The broader collaborative interface forms a loop. A person can correct a claim, revoke a permission, edit a memory, or reject a training trace. Those actions travel upstream and change what later reaches the surface.

The studio is the human pattern behind the environment. I learned to draw in a free art class on Long Island. Everyone received the same figure and instruction. One student made the ears too large. Another gave the figure six fingers. Someone else found a shadow I hadn't seen. Their attempts changed what I noticed and gave me evidence for judging my own drawing. The room was part of the collaboration.

## 2. What this adds

The public Humorphism manifesto proposed translating long-standing collaboration patterns into software [2]. A later essay moved UX from isolated interactions toward the conditions where trust accumulates [3]. This paper adds a test to that program. A translation must reveal which obligations of the human practice survived and which disappeared in software.

Sociotechnical analysis explains how social and technical arrangements jointly produce an outcome [4]. Humorphic design begins with a named human collaboration practice, translates it into software, and tests whether its relational obligations survived. A setting qualifies as a humorphic environment when it names the source practice and its relational obligations. It must also name the software behavior carrying each obligation and the person's recourse when the translation fails.

Three established traditions cover much of this territory. Provenance work traces the evidence, memory, tools, intermediate claims, and actions behind a final answer [5].

Human-agent teamwork research describes mutual predictability, common ground, repair, and directability [6]. Coactive Design turns interdependence into requirements for observability, predictability, and directability [7]. Later work operationalizes those requirements through joint-activity representations and runtime coordination mechanisms [8].

Contestable AI gives people intervention and recourse across time. Human-AI guidelines address remembered interaction, learning from behavior, and cautious updates [9]. Explanatory Debugging lets end users correct a learning system's future predictions [10]. Contestable AI by Design places intervention throughout the system lifecycle [4].

This paper applies those obligations together to the infrastructure of a persistent personal agent. It asks how retrieval selects evidence, memory persists, tools receive authority, and feedback becomes future behavior. The unit of review is one relationship environment derived from a named human practice.

Within that review, directability applies to persistent state transitions. The object is the write path through which an interaction enters memory or training, and the person's later ability to inspect, reverse, or contest that admission.

The argument comes from autobiographical design inquiry [11]. I'm Alicia's builder, participant, evaluator, and category maker. That gives me longitudinal access and concentrates interpretation in one person. Matched controls, raw-run inspection, code review, and adversarial editorial review corrected my first interpretation of the pilot. No external researcher derived the four dimensions. One data-scientist reviewer coded all thirty-three citation chains without model identities or my prior classifications. The coding tests the provenance finding, not the framework's four dimensions.

## 3. Four case-derived dimensions

The provenance failure directly exposed evidence. Its fine-tuning context connected the failure to attention and learning. It contained no consequential action. I therefore treated the incident as the start of a broader audit rather than the source of a complete taxonomy.

That audit followed four existing parts of Alicia: its prompts and retrieval, its evidence paths, its tool permissions, and its preference exports. The resulting dimensions describe this case and the responsibilities examined in this paper. Goal alignment, common ground, coordination, identity, and recovery may require additional dimensions in other systems.

The studio properties and the dimensions operate at different levels. Studio properties are the relational obligations being translated. Attention, evidence, action, and learning are overlapping lenses for auditing how those obligations entered software. Memory, for example, shapes both attention and learning. Correcting evidence can change what becomes salient later.

| Studio obligation | Alicia translation | Audit lens | Status |
| --- | --- | --- | --- |
| Work persists | Shared archive and memory | Evidence, learning | Partial |
| Materials shape perception | Retrieval and attention records | Attention | Partial |
| Tools and the situation answer back | Agent responses and authorized action | Attention, action | Implemented in parts |
| Other attempts train perception | Durable neighboring work | Attention, evidence | Missing |

### Prompts, retrieval, and weights selected what became salient

**Dimension:** Attention

**Human consequence:** The teammate notices on the person's behalf

**Recourse available now:** Memory can be edited if exposed. Weight-level patterns require new data or a model release.

**Actor or access level:** Memory may be participant-facing. System prompts and model releases require builder or model-owner access.

**Artifact and status:** Attention catalogue, proposed from inspectable prompts and memory

### A retrieved synthesis supplied a claim and credited an unopened note

**Dimension:** Evidence

**Human consequence:** The teammate makes claims the person must be able to challenge

**Recourse available now:** The chain can be traced and source files edited manually. Claim invalidation has no dedicated control.

**Actor or access level:** Manual tracing and file editing require builder or file-owner access. Participant-facing invalidation is absent.

**Artifact and status:** Evidence contract, revised after the audit

### Tool permissions changed which initiatives were available

**Dimension:** Action

**Human consequence:** Initiative meets consent

**Recourse available now:** Email can be previewed and explicitly authorized. General revocation and recovery controls are proposed.

**Actor or access level:** Email confirmation is participant-facing. Broader policy changes require builder access.

**Artifact and status:** Consequence map, proposed from implemented permissions

### Reviewed choices entered preference exports

**Dimension:** Learning

**Human consequence:** An interaction may alter the future relationship

**Recourse available now:** A/B wins can be admitted and ties or neither excluded. Revision and removal controls are unimplemented.

**Actor or access level:** Choosing is participant-facing. Export policy, revision, and removal require builder or model-owner access.

**Artifact and status:** Learning gate, partially implemented

The recourse and actor columns distinguish controls available to a participant from powers held by the builder or model owner. A proposed control doesn't satisfy the acceptance criterion until the participant can use it.

## Attention

Attention determines what can become relevant enough to shape a response or action. Alicia's archetypes, contradiction records, lineages, earned questions, weekly reflections, retrieval, and fine-tuning data each make different patterns available.

One weekly note said, “The partnership excels at depth but struggles with peripheral vision, missing his actual movement while serving his stated concerns.” I set down the writing project I kept naming and followed the work my attention had moved toward. The note changed a decision. It doesn't prove whether Alicia sharpened my judgment or replaced part of it.

A prompt-level rule can be opened, named, and changed directly. Once attention moves into weights, the rule must be inferred from behavior and changed through data or a model release. An attention catalogue would record the pattern, its location, its owner, its refusal boundary, and how the person can revise it. Alicia exposes many of those parts without collecting them in one artifact.

### Evidence

Evidence determines what the teammate may present as known and what the person can challenge. A personal archive contains raw notes, quotations, memories, generated syntheses, corrections, and jointly revised work. Retrieval makes them available as text. It doesn't make their authorship or authority equivalent.

The Spanish response exposed the missing distinction. The synthesis was the document Alicia read. The book note was a source the synthesis credited. An evidence contract would preserve both and give the person an action when the chain is wrong.

### Action

Action determines where initiative ends and consent begins. Searching, drafting, editing, sending, publishing, spending, and deleting carry different costs. Knowing how to perform an action doesn't grant authority to perform it.

Alicia can draft an email without approval. Sending it triggers explicit human confirmation. That boundary is implemented. Alicia doesn't yet maintain the proposed consequence map that would record reversibility, reach, authorization, recovery, and who bears the cost of error across all tools.

The practical rule remains simple: let the collaborator rearrange the room and bring back things I didn't request. Protect the exits. Don't let it publish, erase, or speak for me without asking.

### Learning

Learning determines which interactions alter the future relationship. Ordinary use produces abundant feedback and weak evidence about what should become curriculum. I may prefer a response because it's shorter, more flattering, more accurate, or closer to something I already believe.

Alicia Labs exports explicit blind A/B choices as chosen and rejected responses. The Week 35 export holds all ten completed decisions. Weekly exports are cumulative and repeat trial IDs, so concatenating them would duplicate examples. The code excludes ties and neither choices, although the ten decisions contain no example of either. The exclusion path is implemented and remains unexercised in this ledger.

The gate is incomplete. Exported records omit reviewer, admission time, exclusion reason, and corpus version. A training import also needs to deduplicate trial IDs across cumulative exports. A learning gate needs those controls so the person can challenge why one interaction changed the future teammate and another did not.

## 4. An evidence contract in practice

The pilot established four facts. The first comparison appeared meaningful. It was confounded by different runtimes and quantization paths. A matched control erased the apparent model effect. The provenance failure survived.

The reviewer judged that the retrieved intermediary supplied the claim in sixteen relayed citations and part of it in four. Five of twelve direct citations lacked support in the credited target excerpt supplied for review. Across twenty-eight citations with inspected target excerpts, eight had full support, eight partial support, and twelve none.

The labels describe evidence paths. Direct means the cited target was retrieved. Relayed requires an unretrieved target, a visible path through a retrieved intermediary, and at least partial claim supply.

The reviewer labeled twenty of thirty-three bracketed citations Relayed, twelve Direct, and one Unclear. All twenty relayed labels matched the automated path candidates. That match checks the path audit. The reviewer's substantive judgments are in the claim-supply and target-support fields. The protocol's public-packet gate has not been met, so I don't claim independent reproduction. The records also say nothing about how often the failure occurs in other systems or prompts. One shortened title remains unclear. I kept that judgment open rather than turning a plausible path match into a certain one.

Among the twenty relayed citations, credited targets had full support in two, partial support in seven, no support in six, and were not inspected in five. Among the twelve direct citations, six had full support, one partial support, and five none. The one unclear citation's target excerpt had no support. These judgments concern the supplied excerpts, not exhaustive reviews of the underlying works.

The opposite behavior also appeared. One Spanish response began, “Según la síntesis del vault,” before naming the sources that synthesis credited. The intermediary remained visible in prose. The bracket-only audit counted zero citations. A citation count rewarded the flattened chain and missed the preserved one.

The revised evidence contract for the Damasio claim is:

| Field | Record |
| --- | --- |
| Generated claim | Courage requires vulnerability because the body's fear signal must be felt rather than suppressed |
| Claim supplier | Retrieved Alicia synthesis, *Courage and vulnerability are the same gesture* |
| Intermediary author | Alicia-generated synthesis |
| Credited source | Books/Feeling & Knowing |
| Retrieved directly | No |
| Hop count | Two |
| Support inside credited note | Book metadata and a broad bridge, with no passage supporting the specific claim |
| Person's recourse | Trace the chain and edit the synthesis manually today. A dedicated control to invalidate the claim is unimplemented. |

The smallest relevant excerpts make the chain visible:

| Layer | Exact text |
| --- | --- |
| Generated response | “La neurociencia de [[Books/Feeling & Knowing]] (Damasio) añade el matiz crucial: el coraje requiere vulnerabilidad porque exige *sentir* la señal de miedo en el cuerpo, no suprimirla.” |
| Retrieved synthesis | “Damasio's neuroscience clarifies why courage requires vulnerability: true courage is *not* the suppression of the fear-signal but the *integration* of it.” |
| Credited note | “Damasio's neuroscience of feeling connects self-knowledge (mastery) with emotional intelligence in relationships.” |

As an audit artifact, the contract changes review from receipt display to contestable evidence. It tells the reader which document produced the interpretation and whether the credited note can carry the authority assigned to it.

The automated audit now separates bracketed citations from prose attribution. The evidence contract records the claim supplier, credited source, intermediary authorship, retrieval status, hop count, and available correction. A future review should record the claim supplier for every citation, including a directly retrieved one. The visible answer becomes the beginning of review rather than its end.

## 5. Ownership, limits, and the loop

A systems reviewer could describe the four artifacts through familiar mechanisms. Least privilege constrains tools. Data lineage tracks sources. Rebuildable views separate primary material from derived representations. Evaluation separates behavior from inference.

Engineering can verify that a permission is enforced. UX must ask how the boundary changes initiative, consent, and recovery for the person. A provenance system can record a chain. UX must ask whether the person can see and challenge the interpretation that shaped a claim. A training pipeline can version a dataset. UX must ask why an interaction was allowed to change the teammate.

The interface matters because inspection and contest need a human-facing surface. The output is downstream. Correction, consent, and refusal make the relationship a loop. Humorphic design owns both directions.

Alicia remains a single-person case built from a dense private archive. Its training corpus includes earlier Alicia output, and retrieval can feed that voice back into later responses. The pilot establishes no model advantage. Its prompts exposed a broken comparison and a provenance failure. They can't estimate performance.

By this paper's own criterion, Alicia has one clearly participant-facing recourse: email confirmation. Learning exposes a participant-facing choice without later revision or removal. Attention and evidence still depend mainly on builder or file-owner access. Role overlap had made Alicia appear more contestable than it is for an ordinary participant.

The studio translation remains incomplete. Alicia arranges persistent materials and tools, and its behavior can answer back. It has no durable field of neighboring work through which another person's attempts can train perception. That missing obligation appears through the attention and evidence lenses.

The case also doesn't establish that the environment improves human judgment or creativity. The weekly attention scene supports two readings. Alicia may have sharpened my perception or supplied an interpretation I adopted. Distinguishing those outcomes requires longitudinal evidence, free-text reasons, independent coding, and tasks that can reveal strengthened or displaced judgment.

The interface showed a credible answer. The upstream environment concealed how that answer became credible. Design has to own the conditions that produced it and the person's ability to change them.

## 6. Conclusion

Humorphism began by translating partnership into software. Extending it to the environment would make the terms of that partnership reviewable before and after the visible exchange.

The interface is downstream. The relationship is a loop. Humorphic design owns the conditions that produced the encounter and the recourse available to the person inside it.

## Appendix A. Pilot method and controls

The pilot fine-tuned the official Qwen 3.5 9B model [12] with a rank-32 LoRA for eighteen steps. The corpus contained 164 exact-text vault examples: 100 syntheses, 50 earned questions, eleven contradiction analyses, and three lineages. I used 148 examples for training and 16 for validation. “Exact vault text” describes selection and doesn't imply human authorship.

The initial comparison used three frozen prompts. Stock Qwen produced seven completed citations, all to documents retrieved for the prompt. The tuned arm produced seven completed citations and directly grounded two. The comparison couldn't isolate the adapter. Stock ran through Ollama as a larger GGUF. The tuned model ran through MLX as an affine 4-bit artifact. Runtime, quantization, artifact size, and sampling changed together.

The control ran untouched and tuned Qwen as same-size MLX artifacts. Architecture, tokenizer, chat template, sampler, output limit, and retrieved context were identical. A second matched setting recorded equal prompt and model-input hashes, retrieved-source hashes, and stable case-derived seeds between arms. The apparent citation difference disappeared. Citation counts also moved sharply under the second seed. The three prompts diagnose the pipeline and support no performance claim.

The automated audit marked twenty candidate relayed citations across the four matched run records and all three cases. The author manually inspected and classified those twenty. One malformed-link defect also traveled from an agent-generated vault note into retrieval, training, the sole contradiction validation example, and the first audit. Seven of eleven contradiction analyses were malformed. Six entered training.

The independent review protocol covered all thirty-three bracketed citation instances. The reviewer returned a complete, model-blind sheet with twenty relayed, twelve direct, and one unclear label. The unadjudicated cross-tab against the automated candidate set is:

| Automated path audit | Reviewer direct | Reviewer relayed | Reviewer unclear |
|---|---|---|---|
| Candidate relayed (20) | 0 | 20 | 0 |
| Other citations (13) | 12 | 0 | 1 |

This is a comparison with a path audit, not inter-rater agreement. I preserved my earlier item-level judgment for the twenty candidates, but did not lock a complete author sheet for all thirty-three before receiving the reviewer's sheet. That is a protocol deviation. I therefore report no thirty-three-item agreement percentage or kappa. The one unclear title match remains unadjudicated. The original returned workbook has SHA-256 3de58b1a6917f5e668a13af5e5fdce7316300d8fa1fc8f76e6f729a8606a8152.

## Appendix B. Data availability and publication boundary

The private vault and training corpus contain personal and third-party material and can't be released. This paper names only the author's notes, Alicia-generated syntheses, and published works. It contains no identifiable private third-party material.

The audit code and raw run records remain local. The reviewer received a thirty-three-item blinded packet and returned the unadjudicated coding sheet. My earlier twenty positive judgments and the reviewer's twenty relayed labels coincide, but the missing locked author sheet prevents a full agreement calculation. The one unclear item has not been adjudicated with the reviewer.

No public companion package accompanies this preprint. The Damasio trace is inspectable through the excerpts above, while the aggregate cannot be reproduced from public materials. Releasing a redacted packet would require privacy clearance for prompts, run records, code, and citation excerpts. The private vault, full source text, and identifiable third-party material remain outside that boundary.

The technical model and runtime names remain in this academic preprint because they are required to interpret the control. An editorial version for humorphism.com should generalize those details and link to the cleared preprint.